\documentstyle[12pt]{article}

\begin{document}

\begin{titlepage}
\begin{center}

{\Large{\bf On ``Electromagnetic Potential Vectors and Spontaneous Symmetry Breaking"}}\\

\vspace*{1cm}

{\large{\bf V.V.Dvoeglazov}}\\
{\it UAF, Universidad Aut\'onoma de Zacatecas\\
Apartado Postal 636, Zacatecas 98061 Zac., M\'exico\\
E-mail: valeri@fisica.uaz.edu.mx\\
URL: http://fisica.uaz.edu.mx/\~{}valeri/ }
\end{center}

\begin{abstract}
The appearance of terms, which are analogous to ones required for
symmetry breaking, in Lagrangian of Ref.~\cite{Shebalin} is shown
to be caused by gauge invariance of quantum electrodynamics (QED)
and by inaccuracy of the cited author in the choice of canonical variables.
These terms  do not have physical significance within modern quantum electrodynamics.

PACS: 03.50.De, 04.20.Cv, 04.20.Fy, 11.10.Ef .
\end{abstract}

\end{titlepage}

In Ref.~\cite{Shebalin,Shebalin2} the following $k$-- space Lagrangian for
electromagnetic field, interacting with the\,\, current\,\, $\vec j$
and with the charge density $\rho$, has been obtained:
\begin{eqnarray}\label{eq:LagSheb}
\lefteqn{\Lambda(\vec k)=i\Psi^{+}(\vec k)\dot{\Psi}(\vec k)+
\left [\vec A^{*}(\vec k)-\vec k\right ]\vec j(\vec k)-
m\bar\Psi(\vec k)\Psi(\vec k)+}\nonumber\\
&+&i\frac{\vec k\vec C^{*}(\vec k)\rho(\vec k)}{\vec k\,^{2}}-
\frac{\mid\rho(\vec k)\mid^{\,2}}{2\vec k^{\,2}}+\frac{1}{2}
\left (\mid \vec C_{\perp}(\vec k)\mid^{2}-
\vec k^{2}\mid\vec A_{\perp}(\vec k)\mid^{2}\right ).
\end{eqnarray}
The approach was used, in which $\vec A$, the vector potential, and
$\vec C=\dot{\vec A}={\partial\vec A\over \partial t}$ are supposed
to be independent  to
each other. The author of cited paper considers (\ref{eq:LagSheb})
as the Lagrangian with the spontaneous-symmetry-breaking terms
(fourth and fifth in the above formula).

Let us mark, the approach using the additional vector variable
(it is designated as $\vec C$ in Ref.~\cite{Shebalin}), which is different
from field variables, and  it is considered as independent, is not
an innovation.
This is just the well-known Hamiltonian
canonical formalism (see e.g.~\cite{Visconti}-\cite{Kallen}). In Ref.
\cite{Sokolov} the canonical-conjugated variable to $A_{\mu}$ is defined
identically with the quantity $\vec C$ in~\cite{Shebalin}, if we don't
take into account the inessential coefficient ${1 \over 4\pi}$:
\begin{equation}
\pi^{i}=\frac{1}{4\pi}\frac{\partial{\cal L}}{\partial(\partial_t A_i)}=
\frac{1}{4\pi}\frac{\partial\vec A}{\partial t}.
\end{equation}
This canonical-conjugated quantities are due to use of the following Lagrangian:
\begin{equation}\label{eq:lagr}
{\cal L}=-\frac{1}{8\pi}A_{\mu,\nu}A^{\mu,\nu}= -\frac{1}{16\pi}F_{\mu\nu}
F^{\mu\nu}-\frac{1}{8\pi}\frac{\partial A_{\mu}}
{\partial x^{\nu}}\frac{\partial A^{\nu}}{\partial x_{\mu}}
\end{equation}

But, in the case of the $x$-- space Lagrangian
\begin{equation}\label{eq:LagFF}
{\cal L}=-\frac{1}{4}F_{\mu\nu}F^{\mu\nu}
\end{equation}
the quantities
$\vec A$ and $\vec C=\dot{\vec A}$ are
not the canonical-conjugated quantities, as opposed to the case of
classical mechanics where $\vec x$, the coordinate, and $\dot{\vec x}$,
the velocity, are, in fact, the canonical quantities.
It is not clear, what quantzation procedure are implied by the author
of Ref~\cite{Shebalin}.
In the case of canonical quantization the Lagrangian $\sim F_{\mu\nu}F^{\mu\nu}$
does not give us $\pi_0$, which is equal to zero. In the case of Lagrange
quantization it is not clear, what  commutation rules should be implemented,
e.g., for $[\vec A(\vec x),\vec C(\vec{x'})]$. Moreover, in the Lagrange
approach  the field
and the momenta of field are not considered as two independent
quantities. It is also not obvious, how $P_{\mu}$, the energy-momentum
operator, is expressed by $\vec A$ and $\vec C$ in the quantum case.

Let us not forget,
under quantization of electromagnetic field it is impossible to use
the Lorentz condition {\it ab initio}. According to Fermi~\cite{Fermi} it
exists as the condition for the state vectors only. It is necessary to
choose the definite Lagrangian and the definite quantization
approach. and  we are
able to use the Lorentz condition in a weaker form only after setting up the commutation relations:
\begin{equation}
\left (\frac{\partial A^{(-)}}{\partial x}\right )\Phi=0.
\end{equation}

In the case of the Lagrangian (\ref{eq:lagr}) we are able to quantizate
the electromagnetic field canonically, using the variable $\vec\pi=\dot{\vec A}$
as independent. Following the techniques of~\cite{Shebalin}, we then have
the additional terms to the $k$-- space
Lagrangian, which do contract one of the term in (\ref{eq:LagSheb}):
\begin{equation}
{\cal L}^{add}=-\frac{i}{\vec k\,^2}\rho(\vec k)\left (\vec k
\vec C^{*}(\vec k)\right )+\frac{1}{\vec k\,^2}\left (\vec k\vec C(\vec k)
\right )\left (\vec k\vec C^{*}(\vec k)\right )-\left (\vec k\vec A(\vec k)
\right )\left (\vec k\vec A^{*}(\vec k)\right ).
\end{equation}
The total Lagrangian does not contain the symmetry breaking terms.

As a result of gauge invariance of QED it is possible to use
the other Lagrangians differing from (\ref{eq:LagFF}) by the
supplementary term ${1\over\lambda}\left (\frac
{\partial A}{\partial x}\right )^2$, which, on the first view,
brings nothing in (18) of cited paper.
However, in this case the canonical quantities are $\pi_0={1\over \lambda}
({\partial A^{\mu}\over \partial x^{\mu}})$ and $\pi^i=F^{i0}$ and the
expressions (10, 11) between the canonical quantities in
Ref.~\cite{Shebalin} are no longer valid.

Moreover, it is well-known that the Lagrangian can be defined up to
the total derivative only. If we implement the function $\partial^\mu
f_{\mu}=\partial^\mu (g\cdot h)_{\mu}$ it is easy to select $g$ and $h$
in such kind that  both of the symmetry-breaking terms in (18) are
contracted out.

In the end, it is not clear, why the Coulomb gauge ($\vec k\cdot\vec A=0$
and $\vec k\cdot\vec C=0$) was used by the author of~\cite{Shebalin} in the
formula (18), the Lagrangian, but it was not used before, e.g. in (10)
and (15).

In conclusion, the appearance of interaction terms of the form
$a\,\cdot\mid\Psi\mid^{\,^2}+b\,\cdot\mid\Psi\mid^{\,^4}$ in
the QED Lagrangian is caused by gauge invariance of electrodynamics,
implementing the Lorentz condition {\it ab initio}, inaccuracy of
the author in
the choice of the canonical-conjugated quantities. These terms are nonphysical
and can be eliminated
as a result of using the appropriate gauge. Consequently,  they have no
any physical meaning in quantum theory.

I would  still like to mention that investigations of interaction
electromagnetic field with currents deserves  serious
elaboration. In Ref.~\cite{Barut} the equation was presented:
\begin{eqnarray}
\lefteqn{(-i\gamma^\mu \partial_\mu - m)\Psi(\vec x)=}\nonumber\\
&=&e^2\gamma^\mu \Psi(\vec x)\int d\vec y D(\vec x-\vec y)\bar\Psi(\vec y)
\gamma_\mu \Psi(\vec y)
+e\gamma^\mu \Psi(\vec x)A_{\mu}^{in}(\vec x)
\end{eqnarray}
(where $D(\vec x-\vec y)$ is the Green's function for electromagnetic field,
$A^{in}_{\mu}(\vec x)$ is the solution of Maxwell's equations),
which should be resolved (see also~\cite{Kallen}).

\vspace*{5mm}

{\it Acknowledgements.} I am very grateful\, to\, Prof. A. M. Cetto,\, Head\, of the Departamento de
F\'{\i}sica Te\'orica, IFUNAM, for creation of excellent conditions
for research. The technical help of A. Wong is greatly
acknowledged.  
This work has been financially supported by the CONACYT (M\'exico) under
the contract No. 920193.

\vspace*{5mm}

\small{

}
\end{document}